\documentclass{webofc}
\usepackage[varg]{txfonts}   

\usepackage{color}

\newcommand{\gtapprox}{\raisebox{-0.5ex}{$\,\stackrel{>}{\scriptstyle\sim}\,$}}
\newcommand{\ltapprox}{\raisebox{-0.5ex}{$\,\stackrel{<}{\scriptstyle\sim}\,$}}

\begin{document}
\title{Spectroscopy of heavy exotic mesons using lattice QCD static potentials and the Born-Oppenheimer approximation}
%
%

\author{\firstname{}
\lastname{Marc Wagner}\inst{1,2}\fnsep\thanks{\email{mwagner@itp.uni-frankfurt.de}}}

\institute{Goethe-Universit\"at Frankfurt am Main, Institut f\"ur Theoretische Physik, Max-von-Laue-Stra{\ss}e 1, \\ D-60438 Frankfurt am Main, Germany
\and
Helmholtz Research Academy Hesse for FAIR, Campus Riedberg, Max-von-Laue-Stra{\ss}e 12, D-60438 Frankfurt am Main, Germany
          }

\abstract{
I discuss the investigation of heavy exotic mesons using lattice QCD static potentials and the Born-Oppenheimer approximation. I summarize selected recent results for $\bar{b} \bar{b} q q$ tetraquarks, for $I = 0$ bottomonium and for $I = 1$ bottomonium.
}
\maketitle


\section{\label{SEC001}Introduction}

In this talk I summarize theoretical studies of heavy exotic mesons, mostly tetraquarks, based on lattice QCD static potentials and the Born-Oppenheimer approximation. In detail I discuss $\bar{b} \bar{b} q q$ tetraquarks in Sec.~\ref{SEC002} (here and in the following $q \in \{ u , d , s \}$), $\bar{b} b$ ordinary bottomonium together with $\bar{b} b \bar{q} q$ tetraquarks with $I = 0$ in Sec.~\ref{SEC003} and $\bar{b} b \bar{q} q$ tetraquarks with $I = 1$ in Sec.~\ref{SEC004}. $\bar{b} \bar{b} q q$ tetraquarks have not yet been observed experimentally, but one of their charm counterparts, the $T_{cc}$ tetraquark with quantum numbers $I(J^P) = 0(1^+)$, was recently found by LHCb \cite{LHCb:2021vvq}. There are quite a number of $\bar{b} b$ and $\bar{b} b \bar{q} q$ states with $I = 0$ listed in the Review of Particle Physics \cite{ParticleDataGroup:2022pth}. $\bar{b} b \bar{q} q$ tetraquarks with $I = 1$ were also found by experiments, notably the electrically charged tetraquarks $Z_b(10610)$ and $Z_b(10650)$ \cite{ParticleDataGroup:2022pth}.

Due to limited time I do not discuss $\bar{b} b$ hybrid mesons, which can be studied using similar methods. I refer to Refs.\ \cite{Capitani:2018rox,Schlosser:2021wnr} for the lattice computation of hybrid static potentials, to Refs.\ \cite{Braaten:2014qka,Berwein:2015vca,Oncala:2017hop} on how to set up corresponding coupled channel Schr\"odinger equations and to Refs.\ \cite{Bicudo:2018jbb,Muller:2019joq} concerning the lattice computation of hybrid flux tubes.

A more comprehensive recent discussion of both light and heavy tetraquark studies with lattice QCD methods can be found in Ref.\ \cite{Bicudo:2022cqi}.


\section{\label{SEC002}$\bar{b} \bar{b} q q$ tetraquarks}


\subsection{Basic idea of the Born-Oppenheimer approximation}

\begin{figure}[h]
\centering
\begin{center}
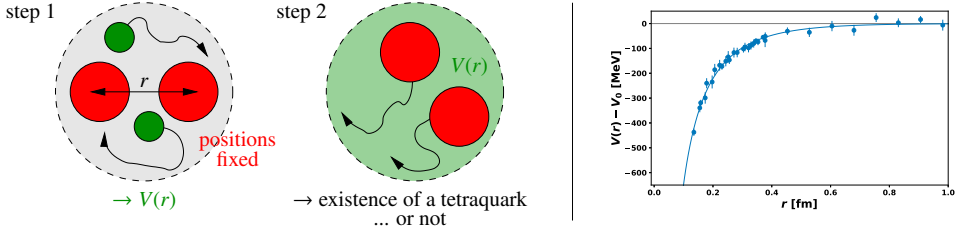
\end{center}
\caption{\label{FIG001}\textbf{(left)}~Study of a $\bar{b} \bar{b} q q$ tetraquark in the Born-Oppenheimer approximation (red circles represent the two heavy antiquarks $\bar{b} \bar{b}$, green circles the two light quarks $q q$). Step~1: Lattice QCD computation of an antistatic-antistatic potential $V(r)$. Step~2: Solving the Schr\"odinger equation with the potential $V(r)$ for the relative coordinate of the $\bar{b}$ quarks provides information about the possible existence of a tetraquark. \textbf{(right)}~The most attractive antistatic-antistatic potential \cite{Lattice2023Mueller}.}
\end{figure}

The basic idea of studying $\bar{b} \bar{b} q q$ tetraquarks with lattice QCD and the Born-Oppenheimer approximation is to split the problem into two separate steps. In the first step lattice QCD is used to compute potentials of two static antiquarks, representing the $\bar{b} \bar{b}$ pair, in the presence of two lighter quarks $q q$. In the second step these potentials are inserted into a Schr\"odinger equation for the relative coordinate of the heavy $\bar b$ quarks. Using standard techniques from quantum mechanics and scattering theory one can then check, whether these potentials are sufficiently attractive to host bound states or resonances, which indicate the existence of QCD-stable tetraquarks or tetraquark resonances. The two steps of the Born-Oppenheimer approximation are sketched in Figure~\ref{FIG001} (left).

\subsection{Computation of antistatic-antistatic potentials with lattice QCD}

To determine antistatic-antistatic potentials $V(r)$, one can use lattice QCD to compute temporal correlation functions
\begin{eqnarray}
\label{EQN012} C(t) = \langle \Omega | \mathcal{O}_{BB}^\dagger(t) \mathcal{O}_{BB}(0) | \Omega \rangle 
\end{eqnarray}
of interpolating operators
\begin{eqnarray}
\label{EQN001} \mathcal{O}_{BB} = (\mathcal{C} \Gamma)_{AB} (\mathcal{C} \tilde{\Gamma})_{CD} \Big(\bar{Q}_C^a(-\mathbf{r}/2)q^a_A(-\mathbf{r}/2)\Big) \Big(\bar{Q}_D^b(+\mathbf{r}/2)q^b_B(+\mathbf{r}/2)\Big) .
\end{eqnarray}
$V(r)$ and the correlation function (\ref{EQN012}) are related via
\begin{eqnarray}
C(t) \propto_{t \rightarrow \infty} e^{-V(r) t} .
\end{eqnarray}
$\bar{Q} \bar{Q}$ are static antiquark operators, $q q$ are light quark operators and $C = \gamma_0 \gamma_2$ is the charge conjugation matrix. The static spin components are coupled with $\tilde{\Gamma} \in \{ (1 + \gamma_0) \gamma_5 , (1 + \gamma_0) \gamma_j \}$, $j = 1,2,3$, which is mostly irrelevant, since energy levels are independent of the orientation of the static spins. The light spin components are coupled with $\Gamma \in \{ (1 + \gamma_0) \gamma_5 , (1 - \gamma_0) \gamma_5\}$, which allows to select the spin and parity quantum numbers characterizing static potentials. Since $q \in \{ u , d , s \}$, one can also select flavor quantum numbers, e.g.\ $qq = ud - du$ corresponds to $I = 0$ or $qq \in \{ uu , ud + du , dd \}$ to $I = 1$. Consequently, there is not just one, but a large number of different potentials. Some of them are attractive, others are repulsive and there are three characteristic asymptotic values for large $\bar Q \bar Q$ separations $r$ corresponding to the energy of two $P = -$ static-light mesons, a $P = -$ and a $P = +$ static-light meson and to two $P = +$ static-light mesons (the $P = +$ static-light meson is around $400 \, \text{MeV}$ heavier than its $P = -$ counterpart). For details I refer to Ref.\ \cite{Bicudo:2015kna}.

The most attractive potential has quantum numbers $(I,|j_z|,P,P_x) = (0,0,+,-)$ (see Ref.\ \cite{Bicudo:2015kna} for details) and corresponds asymptotically to a $B^{(\ast)} B^{(\ast)}$ meson pair, when interpreting the static antiquarks as $\bar b$ quarks (in the static limit the $B$ and the $B^\ast$ meson are mass degenerate). Lattice QCD results from an ongoing computation \cite{Lattice2023Mueller} are shown in Figure~\ref{FIG001} (right).


\subsection{\label{SEC011}Schr\"odinger equation, QCD-stable $\bar{b} \bar{b} u d$ tetraquark with $I(J^P) = 0(1^+)$}

The lattice data points from Figure~\ref{FIG001} (right) can be parameterized consistently by
\begin{eqnarray}
V(r) = -\frac{\alpha}{r} \exp\bigg(-\bigg(\frac{r}{d}\bigg)^p\bigg) + V_0
\end{eqnarray}
with fitting parameters $\alpha$, $d$, $p$ and $V_0$. The $1/r$ term is motivated by 1-gluon exchange at small $r$ and the exponential function reflects color screening at large $r$ and $V_0$ represents the $B^{(\ast)} B^{(\ast)}$ threshold. $V(r)$ can be interpreted as a $\bar b \bar b$ potential and be used in a standard non-relativistic Schr\"odinger equation for the relative coordinate of the heavy $\bar{b}$ quarks,
\begin{eqnarray}
\label{EQN002} \bigg(\frac{1}{m_b} \bigg(-\frac{d^2}{dr^2} + \frac{L (L+1)}{r^2}\bigg) + V(r) - V_0\bigg) R(r) = E R(r) .
\end{eqnarray}
Possibly existing bound states corresponding to energy eigenvalues $E < 0$ indicate QCD-stable $\bar{b} \bar{b} u d$ tetraquarks. There is exactly one bound state for orbital angular momentum $L = 0$ of the $\bar{b} \bar{b}$ pair with binding energy $E = -90_{-36}^{+43} \, \text{MeV}$ with respect to the $B B^\ast$ threshold \cite{Bicudo:2012qt,Bicudo:2015kna}. The quantum numbers $I(J^P) = 0(1^+)$ of the respective QCD-stable tetraquark follow from symmetry arguments. There are no bound states for $L > 0$.


\subsection{Further $\bar{b} \bar{b} q q$ results}


\noindent \textbf{Are there further QCD-stable $\bar b \bar b q q$ tetraquarks with other $I(J^P)$ and/or light flavor quantum numbers?} \\
No, not for $qq = ud$ (neither for $I = 0$ nor for $I = 1$), also not for $qq = ss$ (see Ref.\ \cite{Bicudo:2015vta}). However, $\bar b \bar b u s$ has not yet been investigated within the Born-Oppenheimer approximation and there is strong evidence from full QCD computations that a QCD-stable $\bar b \bar b u s$ tetraquark exists (see e.g.\ Refs.\ \cite{Francis:2016hui,Junnarkar:2018twb,Meinel:2022lzo}).


\vspace{0.2cm}
\noindent \textbf{Effects due to the finite mass of the $\bar b$ quarks and the heavy quark spins} \\
Such effects, which were completely ignored in Sec.~\ref{SEC011}, are expected to be of order \\ $m_{B^\ast} - m_B \approx 45 \, \text{MeV}$. In Ref.\ \cite{Bicudo:2016ooe} they were considered in a crude phenomenological way via a $B B^\ast$ and $B^\ast B^\ast$ coupled channel Schr\"odinger equation with the experimental mass difference $m_{B^\ast} - m_B$ as input. The resulting binding energy is $E = -59_{-30}^{+38} \, \text{MeV}$, i.e.\ significantly reduced with respect to the binding energy obtained from the single channel Schr\"odinger equation (\ref{EQN002}). The physical explanation is that the attractive potential shown in Figure~\ref{FIG001} (right) does not only correspond to a $B B^\ast$ pair, but has also a sizable heavier $B^\ast B^\ast$ contribution.


\vspace{0.2cm}
\noindent \textbf{Are there $\bar b \bar b q q$ tetraquark resonances?} \\
In Ref.\ \cite{Bicudo:2017szl} resonances were studied using standard techniques from scattering theory, however, without considering effects due to the finite mass of the $\bar b$ quarks and the heavy quark spins. Indication for a $\bar b \bar b u d$ tetraquark resonance with $I(J^P) = 0(1^-)$ was found with energy $E = 17_{-4}^{+4}  \, \text{MeV}$ above the $B B$ threshold and decay width $\Gamma = 112_{-103}^{+90} \, \text{MeV}$. In Ref.\ \cite{Hoffmann:2022jdx} the finite mass of the $\bar b$ quarks and the heavy quark spins were included using the approach discussed in the previous paragraph. Within such an improved setup, the $\bar b \bar b u d$ resonance does not exist anymore. The reason is that the relevant attractive potential does not only correspond to a $B B$ pair, but has also a rather large heavier $B^\ast B^\ast$ contribution.


\vspace{0.2cm}
\noindent \textbf{Is the QCD-stable $\bar b \bar b u d$ tetraquark a meson-meson ($B B$) state or a diquark-antidiquark ($D d$) state?} \\
Lattice QCD is the ideal tool to answer this question, because one can use not just one interpolating operator of $B B$ type (Eq.\ (\ref{EQN001})), but also a second interpolating operator of $D d$ type (see Eq.\ (3) in Ref.\ \cite{Bicudo:2021qxj}). One can then compare the contribution of each operator to the antistatic-antistatic potential $V(r)$. For $r \ltapprox 0.2 \, \text{fm}$ clear diquark-antidiquark dominance was found, whereas for $r \gtapprox 0.5 \, \text{fm}$ the system corresponds essentially to two mesons. A simple integration over $r$ leads to an estimate of the composition of the tetraquark, around 60\% $BB$ and around 40\% $Dd$. For details I refer to Ref.\ \cite{Bicudo:2021qxj}.


\section{\label{SEC003}Bottomonium, $I = 0$}

Investigating bottomonium with $I = 0$, i.e.\ $\bar{b} b$ and $\bar{b} b \bar{q} q$ with $\bar{q} q \in \{ \bar{u} u + \bar{d} d , \bar{s} s\}$ is technically more complicated than the previously discussed $\bar{b} \bar{b} q q$ case, because there are two competing channels: a quarkonium channel, $\bar{Q} Q$ (with $Q \equiv b$) and a heavy-light meson-meson channel, $\bar{M} M$ (with $M = \bar{b} q$). At small $r$ the quarkonium channel is energetically lower until at $r \approx 1.0 \, \text{fm}$ the gluonic string breaks and the meson-meson channel is favored. Consequently, a lattice QCD computation of the potentials of both channels is needed, as well as of the mixing potential. The pioneering work reported in Ref.\ \cite{Bali:2005fu} provides these three potentials, but was carried out with rather heavy $u/d$ quark masses ($m_\pi \approx 650 \, \text{MeV}$) and only $2$ dynamical quark flavors. More recent work \cite{Bulava:2019iut} used more realistic light quark masses and $2+1$ dynamical quark flavors, but did not compute the mixing potential.

In Refs.\ \cite{Bicudo:2019ymo,Bicudo:2020qhp,Bicudo:2022ihz} an approach was proposed to crudely adapt the lattice QCD potentials from Ref.\ \cite{Bali:2005fu} to $2+1$ quark flavors and physical quark masses and to use them in a $7 \times 7$ coupled channel Schr\"odiger equation, where the 7 components of the wave function correspond to quarkonium, to the spin-1 triplet of a $\bar B^{(\ast)} B^{(\ast)}$ pair and to another spin-1 triplet of a $\bar B_s^{(\ast)} B_s^{(\ast)}$ pair. After projecting to definite total angular momentum of the light degrees of freedom, one can use standard techniques from scattering theory to determine e.g.\ scattering amplitudes and the $\mbox{T}$ matrix. A numerical search of the $\mbox{T}$ matrix poles in the complex energy plane provides energies of bottomonium bound states and resonances and for the latter also decay widths. Moreover, the components of the resulting wave functions provide the compositions of the states in terms of quarkonium and meson-meson percentages $\% \bar Q Q$ and $\% \bar M M$.

Numerical results are collected in Ref.\ \cite{Bicudo:2022ihz}. Masses of bound states and resonances are consistent with experimentally observed states within the expected large systematic errors (as already mentioned, the lattice QCD results for the potentials were obtained with unphysically heavy $u/d$ quark masses; moreover, heavy quark spin effects and corrections due to the finite $b$ quark mass were neglected). There exist several QCD-stable bottomonium states with clear experimental counterparts. There are also two resonance candidates for the $\Upsilon(10753)$ state recently reported by Belle: an $S$ wave state with $\% \bar Q Q = 1 - \% \bar M M \approx 24 \%$ and a $D$ wave state with $\% \bar Q Q = 1 - \% \bar M M \approx 21 \%$. Moreover, $\Upsilon(10860)$ was confirmed as an $S$ wave state with $\% \bar Q Q = 1 - \% \bar M M \approx 35 \%$.

Note that there is a conceptually similar approach \cite{TarrusCastella:2022rxb} developed independently, which also uses the lattice QCD static potentials from Ref.\ \cite{Bali:2005fu} to study $I = 0$ bottomonium as well as charmonium.

As discussed in Sec.~\ref{SEC002}, static potentials are independent of the heavy quark spins. Thus, systematic errors are possibly large, of order $m_{B^\ast} - m_B \approx 45 \, \text{MeV}$. Such spin effects and further corrections due to the finite $b$ quark mass can be expressed order by order in $1 / m_b$ (see e.g.\ Refs.\ \cite{Eichten:1980mw,Brambilla:2000gk}). The corresponding correlation functions are Wilson loops with field strength insertions. Computations in pure SU(3) lattice gauge theory (i.e.\ without sea quarks) up to order $1 / m_Q^2$ can be found in Ref.\ \cite{Koma:2006fw}. In a subsequent work \cite{Koma:2012bc} these $1/m_Q$ and $1/m_Q^2$ corrections were used to predict low lying (stable) bottomonium states via first order stationary perturbation theory. These corrections led to clear improvements, but convincing and satisfactory agreement with experimental bottomonium results could not be reached.


\section{\label{SEC004}Bottomonium, $I = 1$}

Bottomonium with $I = 1$, i.e.\ $\bar{b} b \bar{q} q$ with $\bar{q} q \in \{ \bar u d , \bar u u - \bar d d , \bar d u \}$, includes the experimentally observed tetraquarks $Z_b(10610)$ and $Z_b(10650)$. Studying this system is technically even more complicated than the previously discussed bottomonium with $I = 0$, because the relevant $\bar{B}^{(\ast)} B^{(\ast)}$ potential does not correspond to the ground state, but to an excited state. The reason is that ordinary bottomonium $\Upsilon \equiv \bar b b$ and a pion (possibly with non-vanishing momentum) have the same quantum numbers, but lower energies. In lattice QCD it is possible to compute excited state energies, but only if all energy levels below are computed as well.

The relevant low-lying potentials were recently computed for the first time in Ref.\ \cite{Prelovsek:2019ywc} (see Figure~2 in that reference). The relevant $\bar{B}^{(\ast)} B^{(\ast)}$ potential is represented by the red data points. For small separations it corresponds to the second excited state ($\Upsilon + \pi$ at rest [blue data points] and with one quantum of momentum [black data points] are below).

Solving a single-channel Schr\"odinger equation with the computed $\bar{B}^{(\ast)} B^{(\ast)}$ potential leads to a bound state close to the $\bar{B}^{(\ast)} B^{(\ast)}$ threshold with binding energy $E = -48_{-108}^{+41} \, \text{MeV}$. This state might be related to $Z_b(10610)$ and $Z_b(10650)$. Clearly, this is an interesting result, however, with a possibly large systematic error. A main source contributing to this systematic error are again heavy spin effects and corrections due to the finite $b$ quark mass, which are currently neglected. Moreover, the coupling of the $\bar{B}^{(\ast)} B^{(\ast)}$ channel to the other channels, in particular to $\Upsilon + \pi$, was ignored. In a follow-up work, three related four-quark sectors with quantum numbers differing in parity and charge conjugation were investigated. Neither of them shows any sign of a bound state \cite{Sadl:2021bme}.


\section*{Acknowledgments}

I thank the organizers of the MESON2023 conference, in particular Stefan Leupold, for the invitation to give this overview talk.
I acknowledge support by the Heisenberg Programme of the Deutsche Forschungsgemeinschaft (DFG, German Research Foundation) -- project number 399217702.



\end{document}